\documentclass[]{aa}
\usepackage{txfonts}
\usepackage[authoryear]{natbib}
\usepackage{amssymb}
\bibpunct{(}{)}{;}{a}{}{,}
\usepackage{epsf}
\usepackage{graphicx}
\newcommand{\zetap}{$\zeta$~Puppis}
\newcommand{\hdn}{HD~96715}
\newcommand{\hdo}{HD~190429A}
\newcommand{\fuse}{\emph{FUSE}}
\newcommand{\iue}{\emph{IUE}}
\newcommand{\copernicus}{\emph{Copernicus}}

\newcommand{\halpha}{\mbox{H$\alpha$}}
\newcommand{\bralpha}{\mbox{Br$\alpha$}}
\newcommand{\brgamma}{\mbox{Br$\gamma$}}
\newcommand{\lyalpha}{\mbox{Ly$\alpha$}}

\newcommand{\htwo}{\mbox{H$_{2}$}}
\newcommand{\lb}{$\lambda$}
\newcommand{\mdot}{$\dot{M}$}

\newcommand{\vsini}{$v\sin\,i$}

\newcommand{\teff}{$T_{\rm eff}$}
\newcommand{\logg}{$\log g$}

\newcommand{\vinf}{$v_{\infty}$}

\newcommand{\rstar}{$R_{*}$}

\newcommand{\tlusty}{{\sc Tlusty}}

\newcommand{\cmfgen}{{\sc CMFGEN}}

\newcommand{\kms}{km\,s$^{-1}$}

\newcommand{\ciii}{\mbox{C~{\sc iii}}}
\newcommand{\civ}{\mbox{C~{\sc iv}}}

\newcommand{\hi}{\mbox{H~{\sc i}}}
\newcommand{\hei}{\mbox{He~{\sc i}}}
\newcommand{\heii}{\mbox{He~{\sc ii}}}
\newcommand{\oiii}{\mbox{O~{\sc iii}}}
\newcommand{\oiv}{\mbox{O~{\sc iv}}}
\newcommand{\ov}{\mbox{O~{\sc v}}}
\newcommand{\ovi}{\mbox{O~{\sc vi}}}

\newcommand{\nii}{\mbox{N~{\sc ii}}}
\newcommand{\niii}{\mbox{N~{\sc iii}}}
\newcommand{\niv}{\mbox{N~{\sc iv}}}
\newcommand{\nv}{\mbox{N~{\sc v}}}

\newcommand{\piv}{\mbox{P~{\sc iv}}}
\newcommand{\pv}{\mbox{P~{\sc v}}}
\newcommand{\pvi}{\mbox{P~{\sc vi}}}

\newcommand{\siiv}{\mbox{Si~{\sc iv}}}
\newcommand{\siv}{\mbox{S~{\sc iv}}}
\newcommand{\sv}{\mbox{S~{\sc v}}}
\newcommand{\svi}{\mbox{S~{\sc vi}}}
\newcommand{\feiv}{\mbox{Fe~{\sc iv}}}
\newcommand{\fev}{\mbox{Fe~{\sc v}}}
\newcommand{\fevi}{\mbox{Fe~{\sc vi}}}
\newcommand{\msolyr}{M$_{\odot}$\,yr$^{-1}$}

\newcommand{\xit}{$\xi_{t}$}

\newcommand{\evi}{$\times\,10^{-6}$}
\newcommand{\evii}{$\times\,10^{-7}$}

\newcommand{\lognHI}{$\log N$(H~{\sc i})}
\newcommand{\lognHII}{$\log N({\rm H}_{2})$}
\newcommand{\logLX}{$\log L_{\rm X}/L_{\rm bol}$}

\begin{document}
   \title{Lower mass loss rates in O-type stars: Spectral signatures of dense clumps
           in the wind of two Galactic O4 stars\thanks{Based 
           on observations made by the NASA-CNES-CSA {\sl Far Ultraviolet
           Spectroscopic Explorer\/} and by the NASA-ESA-SERC {\sl International
           Ultraviolet Explorer\/}, and retrieved from the Multimission Archive at
           the Space Telescope Science Institute (MAST).}  }

   \author{J.-C. Bouret\inst{1}
          \and
           T. Lanz\inst{2}
          \and
           D. J. Hillier\inst{3}
          }

   \offprints{J.-C. Bouret}

   \institute{Laboratoire d'Astrophysique de Marseille, CNRS-Universit\'e de Provence,
              BP 8, 13376 Marseille Cedex 12, France\\
              \email{Jean-Claude.Bouret@oamp.fr}
         \and
     Department of Astronomy, University of Maryland, College Park,
    MD 20742, USA \\
              \email{tlanz@umd.edu}
%
     \and
     Department of Physics and Astronomy, University of Pittsburgh,
    Pittsburgh, PA 15260, USA \\
            \email{hillier@pitt.edu}
    }

   \date{Received: ; Accepted: }

   \abstract{We have analyzed the far-ultraviolet spectrum of two Galactic O4 stars,
   the O4If+ supergiant \hdo\  and the O4V((f)) dwarf \hdn, using archival \fuse\  and
   \iue\  data. We have conducted a quantitative analysis using the two NLTE model
   atmosphere and wind codes, \tlusty\  and \cmfgen, which incorporate a detailed treatment
   of NLTE metal line blanketing. From the far-UV spectrum, we have derived the stellar
   and wind parameters and the surface composition of the two stars. The surface of
   \hdo\  has a composition typical of an evolved O supergiant (nitrogen-rich, carbon
   and oxygen-poor), while \hdn\  exhibits surface nitrogen enhancement similar to
   the enrichment found in SMC O dwarfs which has been attributed to
   rotationally-induced mixing. Following studies
   of Magellanic Cloud O stars, we find that homogeneous wind models could not match the
   observed profile of
   \ov\lb1371 and require very low phosphorus abundance to fit the \pv\lb\lb1118-1128
   resonance lines. We show, on the other hand, that we are able to match the \ov\  and \pv\ 
   lines using clumped wind models. In addition to these lines, we find that \niv\lb1718
   is also sensitive to wind clumping. For both stars, we have calculated clumped wind models
   that match well {\em all} these lines from different species and that remain consistent with
   \halpha\  data. In particular, we have achieved an excellent match of the \pv\
   resonance doublet, indicating that our physical description of clumping is adequate.
   These fits therefore provide a coherent and thus much stronger evidence of
   wind clumping in O stars than earlier claims.
   We show that the success of the clumped wind models
   in matching these lines results from increased recombination in the clumps, hence from
   a better description of the wind ionization structure.
   We find that the wind of these two stars is highly clumped, as expressed
   by very small volume filling factors, namely $f_\infty = 0.04$ for \hdo\ and $f_\infty = 0.02$
   for \hdn. In agreement with our analysis of SMC stars, clumping starts deep in the wind,
   just above the sonic point. The most crucial consequence of our analysis is that the
   mass loss rates of O stars need to be revised downward significantly, by a factor of 3 and more.
   These lower mass loss rates will affect substantially the evolution of massive stars.
   Accounting for wind clumping is essential when determining the
   wind properties of O stars. Our study therefore calls for a fundamental revision in our
   understanding of mass loss and of O-type star stellar winds.

   }

\titlerunning{Clumps in the wind of O-type stars}
\authorrunning{J.-C. Bouret et al.}
   \maketitle


\section{Introduction}
\label{intro_sect} 
The spectrum of hot, massive stars reveals signatures of dense, 
highly-supersonic mass outflows that are driven by the strong
radiation field. These stellar winds have a significant effect on the
evolution of massive stars, as O stars will lose a sizable fraction of
their total mass during their lifetime. Mass loss rates are, therefore, a
crucial parameter of stellar evolution models. These mass loss
rates, like other basic stellar parameters, can be derived from
spectrophotometric analyses based on model stellar atmospheres.
For a recent review of the properties of hot star winds,
see \cite{kudritzki00}.

During the last decade, major advances have been accomplished in
constructing more realistic model atmospheres, which consistently
account for the stellar wind, for departures from the Local Thermodynamic
Equilibrium (LTE), and for metal line blanketing \citep{hubeny95,
hillier98, hauschildt97, pauldrach01, koesterke02, Tueb02}.
Current state-of-the-art non-LTE (NLTE) model atmospheres assume smooth,
homogeneous and stationary winds. There are, however, theoretical
arguments as well as observational evidences that the winds of hot
stars are extensively structured. Radiation hydrodynamics simulations of
the line-driven flow's nonlinear evolution show the line force
to be highly unstable and lead to strong reverse shocks
\citep[see, e.g.,][]{owocki88,owocki94, feldmeier95, owocki99}. We
thus expect the presence of density contrasts in the wind of
O stars, which is supported by several lines of observations.

First, the soft X-ray emission of O stars is widely believed
to originate from shocks propagating through the stellar wind
\citep{lucy82, cassinelli83}. With the advent of high resolution
X-ray spectroscopic capabilities on {\sl Chandra\/} and XMM-{\sl Newton},
detailed wind-shock models can be tested by fitting X-ray line
profiles. The fitted lines indicate that the hot emitting plasma is
located throughout the wind starting close to the photosphere
\citep{kramer03}. However, the high opacity of their wind model
has to be lowered in order to reproduce the X-ray flux level.
\cite{oskinova04} recently offered a solution to this issue
considering highly fragmented winds.

Second, temporal variability of O-star wind line profiles is well
documented and might indicate the propagation of disturbances
throughout the wind, most often observed as Discrete Absorption
Components \citep[see, e.g.,][]{howarth95, massa95}. In the spectrum
of the prototypical O supergiant $\zeta$~Pup, \cite{eversberg98}
observed stochastic variable substructures in \mbox{He~{\sc ii}\lb 4686}
that move away from the line center with time. They explained them as
evidence of blobs or clumps moving outward in the stellar wind.

Third, \cite{hillier91} showed that the weak electron scattering wings
of \mbox{He~{\sc ii}\lb 4686} in the spectrum of Wolf-Rayet (WR) stars
can be explained by clumped wind models and can thus be used to diagnose
density inhomogeneities in hot star winds. The inclusion of clumping
in wind studies of WR stars resulted in lowering their mass loss
rates by a factor of 2 to 4 \citep{moffat94, hamann98, hillier99}.

Lastly, spectral evidence of clumping has been reported in recent studies
of Magellanic Cloud (MC) O stars. \cite{crowther02} and \cite{hillier03} found
that phosphorus needs to be significantly underabundant relative to other
heavy elements in order to match \mbox{P~{\sc v}\lb\lb1118, 1128} in
O supergiants. Alternatively, theoretical line profiles from clumped wind models match
the observed \mbox{P~{\sc v}} lines, assuming a ``standard'' phosphorus abundance.
>From studying the wind ionization
of a large sample of LMC O stars, \cite{massa03} came to a similar conclusion
about the \mbox{P~{\sc v}~} resonance lines providing an indication
of wind clumping. Moreover, \cite{hillier03} showed that the clumped wind models
match strong Fe lines, while homogeneous wind models predict marked blue
asymmetries that are not observed.
Independently, \cite{bouret03}
successfully matched for the first time the \mbox{O~{\sc v}\lb1371} line
profile using clumped models. A good match to this line was never achieved
with homogeneous wind models. Bouret et al. found that clumping starts just
above the photosphere. They derived mass loss rates for several O dwarfs
in the SMC cluster NGC~346, which are a factor of 3 to 10 smaller
than those obtained from smooth winds. This reduced mass loss is expected
to significantly alter the predicted evolution of these stars.

\cite{puls96} showed the existence of a tight correlation between the modified
wind momentum and the stellar luminosity. This Wind momentum -- Luminosity
Relation (WLR) varies for Galactic and MC O stars because
of the different metallicities. Stars with lower abundances of heavy elements
have weaker winds \citep{walbornSMC}. Mass loss rates predicted using 
radiative line-driven wind theory bring support to the WLR
\citep{vink00}, hence allowing a derivation of O star luminosities and
of extragalactic distances. However, the dependence of the WLR with the
luminosity class remains an open problem. Recent theoretical studies do not
predict such a dependence while, empirically, a difference in the WLR for supergiants
and for dwarfs has been found. \cite{repolust04} argued that this discrepancy
between the empirical and theoretical WLR's might be solved if the mass loss
rates derived from H$\alpha$ are affected by clumping in the lower wind region.
Clumping would result in decreasing the mass loss rates of Galactic O stars with
H$\alpha$ emission by a factor of about $\sqrt 5$.

Despite mounting evidence that hot star winds do not appear to be smooth and
homogeneous, most studies still assume so with the hope of determining average
wind properties at least. In MC O stars, however, spectral analyses accounting for
clumping yield significantly lower mass loss rates, with potentially crucial
consequences for our understanding of massive star evolution,
and for the radiative line-driven wind theory and its
application to extragalactic studies using the WLR.
In this context, we believe that it is of special importance to extend the recent
studies of MC O stars to Galactic O stars which have stronger winds. Furthermore,
it is essential to show that a consistent picture may be drawn from the different
diagnoses of clumping, in particular \mbox{P~{\sc v}\lb\lb1118, 1128},
\mbox{O~{\sc v}\lb1371}, and H$\alpha$, ensuring that the spectral diagnoses of
wind clumping are not spuriously affected by abundance or ionization effects. 

\begin{table*}[]
\centering \caption{Basic stellar data, \fuse\  datasets, and interstellar column densities.}
\begin{tabular}{lccccccccc}
\hline
\hline
  Star       & RA         &     Dec      &Spectral  & \fuse\   & $V$    & $B-V$ & $E(B-V)$ & \lognHI     & \lognHII \\
             & (J2000)    &    (J2000)   &Type      &   ID     &  (mag) &(mag)  & (mag)    & (cm$^{-2}$) & (cm$^{-2}$) \\
\hline
  HD 96715   & 11 07 32.9 & -59 57 49    &O4V((f))  & P1024301 &  8.27  & 0.10  & 0.42     & 21.3        & 19.7 \\
  HD 190429A & 20 03 29.4 & +36 01 30    &O4If+     & P1028401 &  7.12  & 0.14  & 0.51     & 21.5        & 20.3   \\
\hline
\end{tabular}
\label{tabone}
\end{table*}

In this paper, we investigate the properties of two Galactic O4 stars, \hdo\ and \hdn,
in order to derive their properties and address these issues. These two
stars were chosen to cover a wide range of luminosity classes, thus investigating
wind clumping in O dwarfs and in O supergiants, and to compare Galactic stars to SMC
O4V stars for which Bouret et al. found evidence of clumping. The general stellar
properties, along with the observational data and reduction are presented in
\S\ref{obs_sect}.~~Sect.~\ref{modass_sect} considers our
model atmospheres. We discuss the methodology
used to derive the stellar parameters, chemical abundances, and
wind parameters, and the related uncertainties, for each star, in
\S\ref{specan_sect}. Our results are then put into a broader
context in \S\ref{disc_sect}, comparing in
particular the derived mass loss rates to other predictions based
on homogeneous wind models. General conclusions are
summarized in \S\ref{concl_sect}.


\section{Stellar sample and observations}
\label{obs_sect}

We have selected two Galactic O4 stars, \hdo\ and \hdn, which have been
previously observed in the far ultraviolet (FUV) by \fuse\  and \iue,
to investigate the FUV spectral signatures of clumping 
(\mbox{O~{\sc v}}, \mbox{P~{\sc v}}). The two stars are hot enough
(O4 spectral type) to reveal \mbox{O~{\sc v}\lb1371} in their spectrum. We have
selected one supergiant and one dwarf providing an initial comparison
between the two classes. The two stars are part of the \fuse\  atlas
of Galactic O stars \citep{pellerin02}, from which we extracted
basic stellar data (Table~\ref{tabone}). 
\hdo\  is a bright O
supergiant for which extensive observational and modeling work has been published
\citep[e.g.,][]{walborn00, markova04, garcia04}. 
\cite{garcia04} recently reported the first detailed analysis of the O4V((f)) star
\hdn.

We have extracted \iue\  short wavelength, high resolution spectra from
the Multimission Archive at the Space Telescope Science Institute (MAST).
The SWP spectra cover the spectral range, \lb\lb1150-2000\,\AA,
at a resolving power $R = 10,000$. Each spectrum was processed with the
NEWSIPS package. We have selected all spectra obtained through
the large aperture, that is, 16 spectra for \hdo\  and 4 spectra
for \hdn\ (Table~\ref{SWPTab}). 
These spectra do not show conspicuous signs of variability, in flux as well as 
in line profiles. Although we cannot rule out definitively variability (in particular
for the supergiant \hdo), this absence of variations justifies our co-adding
all merged extracted spectra so as to form an average spectrum for each star.
Data points flagged by
the NEWSIPS software have been excluded; in particular, this concerns
the saturated portion of the long-exposed images SWP\,43980 and SWP\,43981 at
$\lambda\ga 1700$\,\AA. Finally, we smoothed the co-added spectra to
a resolution of 40~\kms\  in order to
increase the signal-to-noise ratio. The spectra of both stars show a large
number of narrow lines of interstellar (IS) origin. This IS contamination
becomes an issue especially for \hdn\ because its apparent rotational
velocity is relatively low \citep[\vsini = 80 \kms;][]{howarth89}.

The processed \fuse\ spectra have been retrieved from MAST too. 
The nominal spectral resolution is 20,000, or about 20~\kms.
Details about the observations and data reduction can be found in
the \fuse\  atlas of Galactic OB spectra presented by \cite{pellerin02}. 
Individual sub-exposures have been co-added for each segment and
then merged to form a single spectrum, using Lindler's FUSE-REGISTER program.
We avoided contamination by the so-called worm artefact \citep{sahnow00}
by using only the LiF2A spectra on the long-wavelength side (\lb\lb1086-1183\,\AA)
of the spectrum. Finally, the co-added merged spectra have been smoothed
to a 30~\kms\  resolution to enhance the signal-to-noise ratio.

>From the \iue\ and \fuse\ spectra, we have measured the atomic and
molecular hydrogen column densities towards the two stars,
fitting \lyalpha\ and H$_{2}$ lines, respectively.
The values are listed in the rightmost columns of Table~\ref{tabone},
indicating that the \fuse\ spectra suffer from severe blending
from the broad absorption bands of \htwo. This absorption especially becomes
a serious issue shortwards of 1006\,\AA\  (SiC2A channel), and thus
little information about the stars can be inferred from the analysis
of this part of the spectra. Furthermore, numerous IS metal lines are
also present in the \fuse\ spectra, superimposed to the stellar lines.
These lines are narrower than the stellar lines but, because of
their large number \citep[see Table~3,][]{pellerin02}, they
may complicate the analysis of the stellar components (especially for
weak lines in \hdn).

\begin{table}[b]
\centering \caption{\iue\  SWP spectra.}
\begin{tabular}{llcl}
\hline
\hline
  Star       & Image     & $t_{\rm exp}$ & Obs. Date \\
             &           &   (s)         &            \\
\hline
HD 96715     & SWP 21999 & 3000          &  1984-01-13 \\           
             & SWP 22000 & 1800          &  1984-01-13 \\           
             & SWP 43980 & 9000          &  1992-02-13 \\           
             & SWP 43981 & 9000          &  1992-02-13 \\ [2mm]
HD 190429A   & SWP 04903 & 1515          &  1979-04-09 \\           
             & SWP 38958 & 1500          &  1990-06-02 \\                                  
             & SWP 38965 & 1500          &  1990-06-02 \\                                  
             & SWP 38970 & 1500          &  1990-06-03 \\                                  
             & SWP 38973 & 1500          &  1990-06-03 \\                                  
             & SWP 38978 & 1500          &  1990-06-03 \\                                  
             & SWP 38981 & 1500          &  1990-06-03 \\                                  
             & SWP 38986 & 1500          &  1990-06-03 \\                                  
             & SWP 38989 & 1500          &  1990-06-03 \\                                  
             & SWP 38994 & 1500          &  1990-06-04 \\                                  
             & SWP 38998 & 1500          &  1990-06-04 \\                                  
             & SWP 39003 & 1500          &  1990-06-04 \\                                  
             & SWP 39006 & 1500          &  1990-06-04 \\                                  
             & SWP 39011 & 1500          &  1990-06-04 \\                                  
             & SWP 54573 & 1500          &  1995-05-02 \\                                  
             & SWP 55986 & 1500          &  1995-09-22 \\                                  
\hline
\end{tabular}
\label{SWPTab}
\end{table}


\section{Modeling assumptions}
\label{modass_sect} 

Unified models, with a consistent treatment of the photosphere and the
wind, are mandatory to analyze the P~Cygni profile of strong lines
in the UV spectra of O stars and, thus, to derive the basic wind
parameters (mass loss rate, terminal velocity). However, the bulk of
spectral lines in O stars are formed in the photosphere where
velocities are small and geometrical extension is negligible.
Hydrostatic, fully-blanketed, NLTE photospheric models may then
remain a preferable alternative to unified models with a
simplified treatment of metal line blanketing. We demonstrated
this point in the case of low metallicity stars in the SMC, where
stellar winds are weaker than those of Galactic stars \citep{bouret03}.
Here, we deal with Galactic objects and, in particular, with an
extreme O4If+ supergiant (\hdo) which exhibits a strong stellar wind.
This is thus an exemplar case for studying the wind contribution
to weak photospheric lines, comparing the predictions of photospheric
and unified models to observations. We have performed such an
analysis using model atmospheres calculated with the photospheric
program, \tlusty\  \citep{hubeny95}, and with the unified model code,
\cmfgen\  \citep{hillier98}. The
atomic data used by the two codes mostly come from the same sources,
thus ensuring a meaningful comparison of the stellar parameters derived
by the two codes. \cmfgen\  does not solve the
full hydrodynamics, but rather assumes the density structure. We use
a hydrostatic density structure computed with \tlusty\  in the deep
layers, and the wind part is described with a standard $\beta$-velocity
law. The two parts are connected below the sonic point at
$v(r)\approx 15$\,\kms. For more details on these
two codes, we refer to \cite{hubeny95}, \cite{lanz03}, \cite{hillier98},
\cite{hillier03}, and \cite{bouret03}.  

Radiatively driven winds are intrinsically subject to
instabilities, resulting in the formation of the discrete
structures or ``clumps''. As discussed in \S1, there are
several observational evidences as well as theoretical
arguments that foster the concept of highly-structured winds.
To investigate spectral signatures of clumping in the winds of \hdn\ and
\hdo, and its insuing consequences on the derived wind parameters, we
have thus constructed clumped wind models with CMFGEN. A simple,
parametric treatment of wind clumping is implemented in CMFGEN,
which is expressed by a volume filling factor, $f$, and which assumes
a void interclump medium and the clumps to be small compared to the photons
mean free path. The filling factor is such that $\bar{\rho} = f\rho$,
where $\bar{\rho}$ is the homogeneous (unclumped) wind density.
The filling factor decreases
exponentially with increasing radius (or, equivalently, with increasing
velocity)
\begin{equation}
f = f_\infty + (1-f_\infty) \exp(-v/v_{\rm cl}),
\end{equation}
where $v_{\rm cl}$ is the velocity at which clumping starts. \cite{hillier03}
and \cite{bouret03} found that clumping starts close to the photosphere.
We have thus adopted, $v_{\rm cl}=30$\,\kms, that is, clumps start
forming just above the sonic point.

The observation of strong resonance lines of highly-ionized species,
like \mbox{O~{\sc vi}}\lb\lb1032, 1038 and \mbox{N~{\sc v}}\lb\lb1238, 1242,
provides an indirect evidence of X-ray emission in the stellar winds. 
An important consequence of the X-ray and EUV shock radiation is
to enhance photoionization that results in ``wind superionization''.
Auger processes are accounted for in calculating the wind ionization \citep{cassinelli79}.
X-rays probably originate from the cooling
zone of post-shock regions, while shocks arise from radiative
instabilities inherent to line-driven winds \citep{lucy70, lucy82,
owocki88}. X-ray emission has been observed in O-type stars
of all luminosity classes \citep{chlebo91}. To reproduce the
\mbox{O~{\sc vi}} and the \mbox{N~{\sc v}} lines, we have accounted
for shock-generated X-ray emission in the final modeling stage.
Because the two selected stars have not been observed with X-ray
satellites, we have assumed the luminosity ratios, \logLX, measured for stars
of same spectral subtype and luminosity class. For \hdo, we have adopted \logLX\ = -7.1,
quoted by \cite{pauldrach01} for the O4~I(n)f star \zetap, whose spectral
and physical properties are expected to be quite similar to those
of \hdo; for \hdn , we have adopted \logLX\ = -6.6, measured
for the O4V((f)) star HD~46223 \citep{chlebo91}.


\section{Spectral analysis}
\label{specan_sect}
Following the methodology outlined in \cite{bouret03}, we
determined initial estimates of the stellar parameters with \tlusty\
model atmospheres \citep{lanz03}. Subsequent analysis is performed
with \cmfgen, using \tlusty\  parameters as initial inputs. At this
stage, we still allowed for changes in the photospheric parameters,
taking advantage of the realistic description of photospheric layers
by \cmfgen. We always checked for consistency between \tlusty\ 
and \cmfgen\  determinations.

Hydrogen and helium lines in the optical spectrum are the classical
diagnoses used in spectral analyses of O stars. In this paper, we
present an analysis based on the far-ultraviolet spectrum exclusively.
\cite{heap05} discussed in detail the UV lines and ionization equilibria
that are good indicators of stellar parameters (effective temperature,
surface gravity). Table~\ref{LineTab} lists the most important lines
used in our analysis and the quantities they are most sensitive to.

As a reference, we adopted the solar abundances from \cite{grevesse98}. However,
we note that significant downward revisions of the abundance of light elements
in the solar photosphere have been recently proposed
on the basis of a 3-D hydrodynamical model of the solar atmosphere
\citep{asplund04}, bringing the latter values in better agreement with
surface abundances of B-type stars in the solar neighborhood. In this paper,
all chemical abundances are quoted by number density relative to hydrogen,
or relative to \cite{grevesse98} solar values.

\begin{table}
\centering \caption{Main spectral lines used in the analysis of the photospheric
and wind properties of \hdo\ and \hdn, and their sensitivity to stellar parameters.}
\label{LineTab}
\begin{tabular}{lccc} \hline \hline 
 & & \multicolumn{2}{c}{Dependencies} \\
\cline {3-4}
 Line   & \lb\ (\AA)             &  \hdo           &      \hdn  \\   
\hline       
 \siv\   & 1073-99              & \teff, \mdot, S/H   & $\ldots$ \\
 \pv\   & 1118-28              & \mdot, $f$, P/H   & \xit, P/H \\
 \oiii\ & 1150-54         & \xit, O/H         & \xit, O/H       \\
 \civ\  & 1169              & \teff, \mdot, C/H & \teff, \xit, C/H \\
 \ciii\ & 1176                   & \teff, \mdot, C/H & \teff, \xit, C/H  \\
 \niii\ & 1182-84              & \xit, N/H         & \xit, N/H     \\
 \fevi\ & 1250-1350              & \teff, \xit, Fe/H & \teff, \xit, Fe/H     \\
 \oiv   & 1338-43             & \teff, \mdot, O/H & \teff, \mdot, O/H \\
 \ov\   & 1371			         & \teff, \mdot, $f$, O/H    & \teff, \mdot, $f$, O/H \\
 \fev\  & 1350-1500		         & \teff, \xit, Fe/H & \teff, \xit, Fe/H     \\
 \siiv\ & 1393,1402              & \teff, \mdot, $f$ &  $\ldots$                 \\ 
 \ciii\ & 1426-28              & $\ldots$		     & C/H, \teff     \\
 \sv\   & 1502                   & \mdot, $f$ , S/H  & \xit, \teff \\ 
 \civ\  & 1548-50		         & \vinf, $\beta$, \xit   & \vinf, $\beta$, \xit \\
 \feiv\ & 1500-1750              & \teff, \xit, Fe/H & \teff, \xit, Fe/H \\
 \heii\ & 1640                   & \mdot, $f$, He/H  & \mdot, He/H \\
 \niv   & 1718   		         & \mdot, $f$, N/H 	         & \mdot, $f$, N/H 	 \\
 \niii\ & 1748-52              & \xit, N/H         & \xit, N/H     \\ 
 \civ\  & 1860                   &  C/H, \teff       & $\ldots$             \\
 \ciii\ & 1875-78              &  C/H, \teff       & $\ldots$             \\ 
 \niii\ & 1885                   & \xit, N/H         & $\ldots$        \\ 
\hline       
\end{tabular}
\end{table}

\subsection{\hdo\ - {\rm O4~If+}}
\label{hdo_sect}
The \iue\  and \fuse\  spectra offer a broad range of ionization stages
and diversity of species that may be used to constrain the effective temperature:
\heii, \ciii, \civ, \niii, \niv, \siv, \sv, \svi, \feiv, \fev, \fevi. 
Following \cite{heap05}, we used in priority ratios between
successive ions of C, N, O and Fe. Both the \ciii\lb1176/\civ\lb1169 and the
\feiv/\fev \footnote{\feiv\ lines between 1500 and 1700\,\AA, and
\fev\ lines between 1300 and 1500\,\AA.} line ratios indicate \teff $\la 40,000$\,K.
The \fevi\ lines in the 1250-1350\,\AA\  range
are quite weak at these temperatures. Moreover, the
\niii\lb\lb981-991 lines are found to be very sensitive to \teff,  as
noticed by \cite{crowther02}, and they set a lower limit \teff\ $\ga 37,500$\,K.
The ionization ratio, \oiv\lb\lb1338-1343/\ov\lb1371, further supports
this lower limit on \teff. On the other hand, \siiv\lb\lb1393-1402, which
are also very sensitive to temperature, suggest \teff\ $\approx 35,000$\,K.
It is unlikely, however, that a contamination of the \iue\  spectrum
by the companion is responsible for this lower value. Indeed, the large \iue\ 
aperture contains both the primary and the companion HD~190429B
(at~$2\arcsec$), but the O9.5~II spectral type of the companion \citep{walborn00}
implies that its contribution remains small (the luminosity ratio should
be about 5). Overall, we found that the best fit
is achieved for \teff = 39,000\,K, with an uncertainty better than 1000\,K.
This value is in excellent agreement with the results from \cite{markova04},
who used a calibration of \teff\ with the spectral type based on results
obtained from the analysis of hydrogen and helium lines in the
optical spectrum \citep{repolust04}. 
On the other hand, \cite{garcia04} derived a slightly lower value, \teff = 37,500\,K,
based on UV lines formed in the wind.
These lines are sensitive to the detail of the wind ionization, hence to
wind clumping that Garcia~\& Bianchi neglected. Moreover, they used carbon and oxygen lines
assuming solar abundances while admitting that these two species might have
highly non-solar abundances (see below).

Our initial estimate of the surface gravity comes from the relation between
the spectral type and \logg\ for luminosity class~I stars
\citep[Fig.~2,][]{markova04}. This relation yields \logg\ = 3.65
for an O4 supergiant. We then explored models with \logg\ ranging
from 3.5 to 3.75 (by steps of 0.05\,dex, for \teff\ fixed at 40,000\,K).
We found that the shape of the observed spectral energy
distribution is indeed best reproduced for $\log g\approx 3.6 - 3.65$
\cite[see][ for the dependence of the SED as a function of \logg,
at a given \teff]{lanz03}. We have adopted, \logg\ = 3.6, for the rest of
the analysis.

The visual photometry listed in Table~\ref{tabone} pertains to both
components A+B. Based on the Hipparcos magnitude difference of the
two components, \cite{walborn00} obtained $V = 7.12$ for component~A
alone. The absolute visual magnitude is then derived, assuming
a distance $d = 2.29$ kpc for the CygOB3 association \citep{humphreys78}.
The stellar luminosity is calculated by applying a bolometric correction from
\cite{lanz03}. The derivation of the stellar radius then follows
straightforwardly.

The microturbulent velocity, \xit, is determined from the iron
line strengths (\feiv\ and \fev). The iron abundance is kept to
the solar value throughout this analysis. As noted
in \cite{hillier03} and in \cite{bouret03}, \oiv\lb\lb1338-1343
is also sensitive to \xit.
However, the oxygen abundance may have a non-solar 
value at the surface of an O4~If+ star (see below). A consistent fit
to the aforementioned lines is obtained for \xit\ = 15\,\kms.
We note that these lines have been similarly used in other studies \citep{bouret03, heap05},
yielding a range of microturbulence values for O stars (\xit\ = 2--25\,\kms).
It is therefore unlikely that the high microturbulence derived here results from
inadequate model atoms.

 \begin{figure*}[]
   \centering
   \rotatebox{0}{\includegraphics[width=18cm]{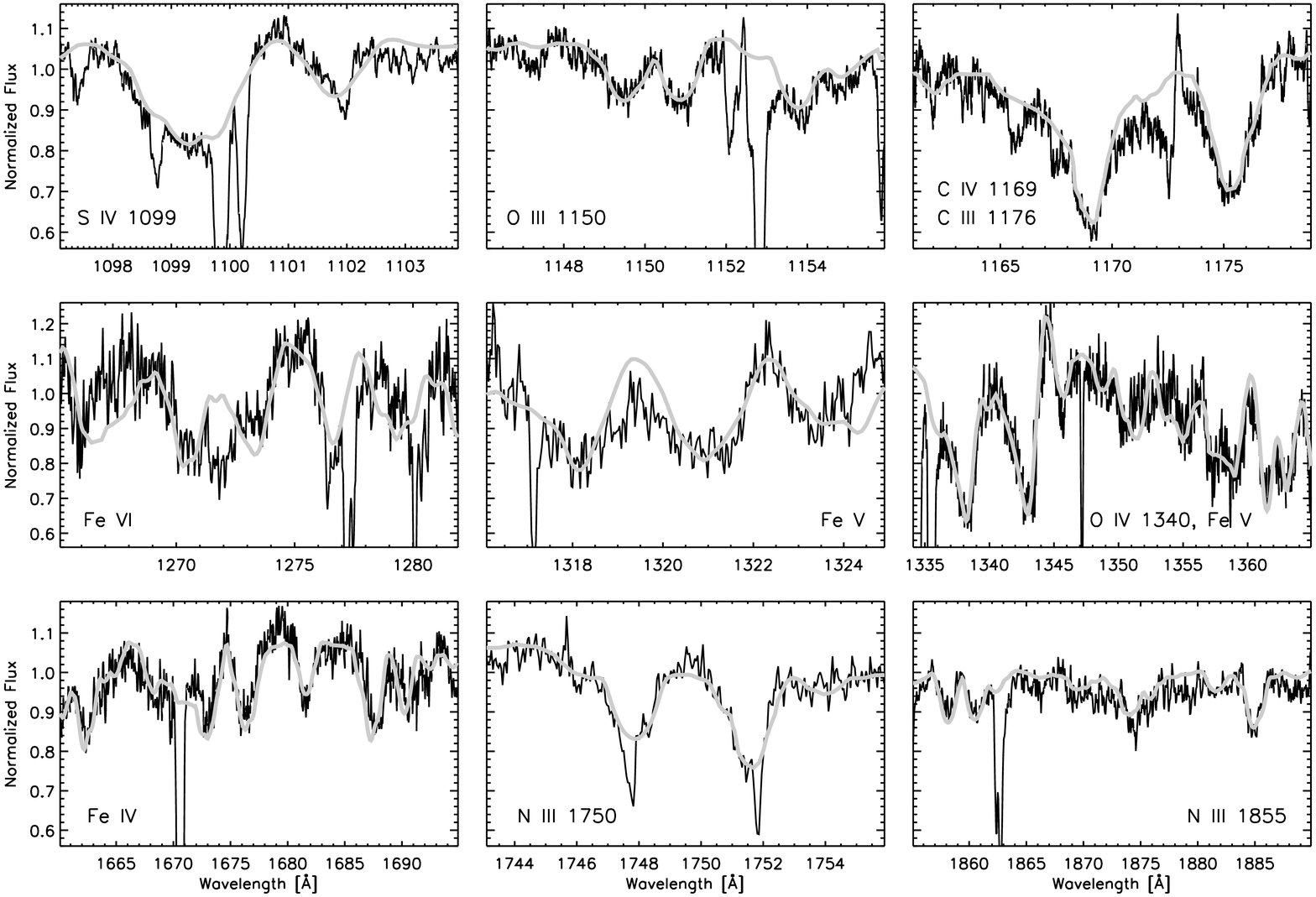}}
      \caption[12cm]{Best fit to the photospheric lines used to derive the 
      stellar parameters of \hdo\ (\teff = 39,000\,K, \logg\ = 3.6, 
      \xit\ = 15\,\kms). The adopted abundances are listed in Table~\ref{TabRes}}.
         \label{figone}
   \end{figure*}

The helium abundance is poorly constrained by the FUV spectrum alone.
We first used $y$ = He/H = 0.1 (by number), but
subsequently  we increased the abundance to $y$ = 0.2. The higher
helium abundance provides a better fit to
\heii\lb1640 (once the mass loss rate is fixed at the value
derived from other lines). The photospheric \heii\lb1085 line,
which is not affected by the wind in our models, could not be used
because of strong blends from IS \nii\  lines. The
enhanced helium abundance is consistent with the evolutionary
status of an O4 supergiant \citep{meynet00}, but we
cannot rule out that it may be further revised when optical
data is analyzed in addition to the FUV spectrum.
A simultaneous fit of all \hei\ and \heii\ 
lines throughout the spectrum is, however, rarely (if ever) achieved
\cite[see, e.g.,][]{hillier03}. Additionally, we note that
changes in the helium abundance impact the predicted
\pv\lb\lb1118-1128 lines, because the ionization threshold of \pv\ 
is at 191\,\AA, close to the \heii\  Lyman limit. Models with
$y$ = 0.2 yield a phosphorus abundance that is closer to the
solar value compared to values derived with $y = 0.1$ models. On the basis
of these various arguments, we finally adopted $y$ = 0.2.

The carbon abundance is constrained by the \ciii\lb1176,
\ciii\lb1923, and \civ\lb1169 photospheric lines.
Only models strongly depleted in carbon match these lines.
A low carbon abundance is also consistent with the 
\ciii\lb977 and the \civ\lb\lb1548-1550 resonance
lines, although these lines are not sensitive indicators
because they show saturated P~Cygni profiles that are
only weakly sensitive to the wind velocity law.
We also checked that the models do not predict a
strong \ciii\lb4647-51 emission that is not observed in
the optical spectrum \citep{walborn00}. The best fit to the
lines discussed above is obtained for C/C$_\odot$ = 0.05.
This value is consistent with the evolved nature of \hdo.

All nitrogen FUV lines, but \niv\lb1718, show that the surface of
\hdo\ has been enriched in nitrogen. This conclusion is additionally
supported by the observed \niii\lb\lb4634-4640 emission
\citep[see, e. g.,][]{walborn00}.
We have derived a nitrogen surface abundance, N/N$_\odot$ = 4.0.
The nitrogen enrichment of HD 190429A is consistent with its being highly 
evolved, as significant nitrogen enrichment together with carbon depletion 
is predicted by stellar evolution models for stars in the helium burning phase.
Note in this respect that
\cite{conti95} and \cite{walborn00} already suggested an advanced
evolutionary status for \hdo\, based on K-band and optical data
(respectively), which they interpreted as indications that the
star is well advanced toward the WN stage.

The lack of wind-insensitive oxygen lines hampers an accurate
determination of the oxygen abundance. The strong
\ovi\lb\lb1032-1038 are fully controlled by the X-ray flux
and other wind parameters. The \ov\lb1371 line is rather
weak and is most sensitive to the mass loss rate and clumping
parameters. Finally, although mostly formed in the photosphere, the
\oiv\lb\lb1338-1343 lines show a blue asymmetry that indicates
a contribution from the wind. Assuming \xit\ = 15\,\kms\ (see above),
we had to adopt a low oxygen abundance, O/O$_\odot$ = 0.1, in order to
match the \oiv\lb\lb1338-1343 lines. However, we always failed to
reproduce the \ov\lb1371 line with homogeneous wind models.
This low abundance is also supported by the weak \oiii\ triplet
(\lb\lb1150-1154\,\AA) in the \fuse\ spectrum.

Phosphorus and sulfur abundances are not expected to be affected
by nucleosynthetic processes. We thus adopted initially (and kept for
most of the analysis) solar abundances. There are, however, indications
that both elements might be slightly depleted. For homogeneous wind models,
a low phosphorus abundance, P/P$_\odot$ = 0.1, is required to match
the \pv\lb\lb1118-1128 resonance doublet. On the other hand, we 
derive P/P$_\odot$ = 0.5 from a clumped model
with a filling factor $f_\infty = 0.04$ (see below). All sulfur ions
indicate that S/S$_\odot$ = 0.5 for homogeneous models and
S/S$_\odot$ = 0.9 for clumped models ($f_\infty = 0.04$). 
Similarly, \cite{hillier03} derived lower than one-fifth solar-scaled P and 
S abundances for the SMC O7~Iaf+ supergiant AV~83. However,
given the recent downward revision of the solar P and S abundances \citep{asplund04},
we argue that the abundances derived from the clumped wind model roughly
remain in agreement with solar values within the uncertainties (0.1 to 0.2\,dex).

 \begin{figure*}[]
   \centering
   \rotatebox{0}{\includegraphics[width=18cm]{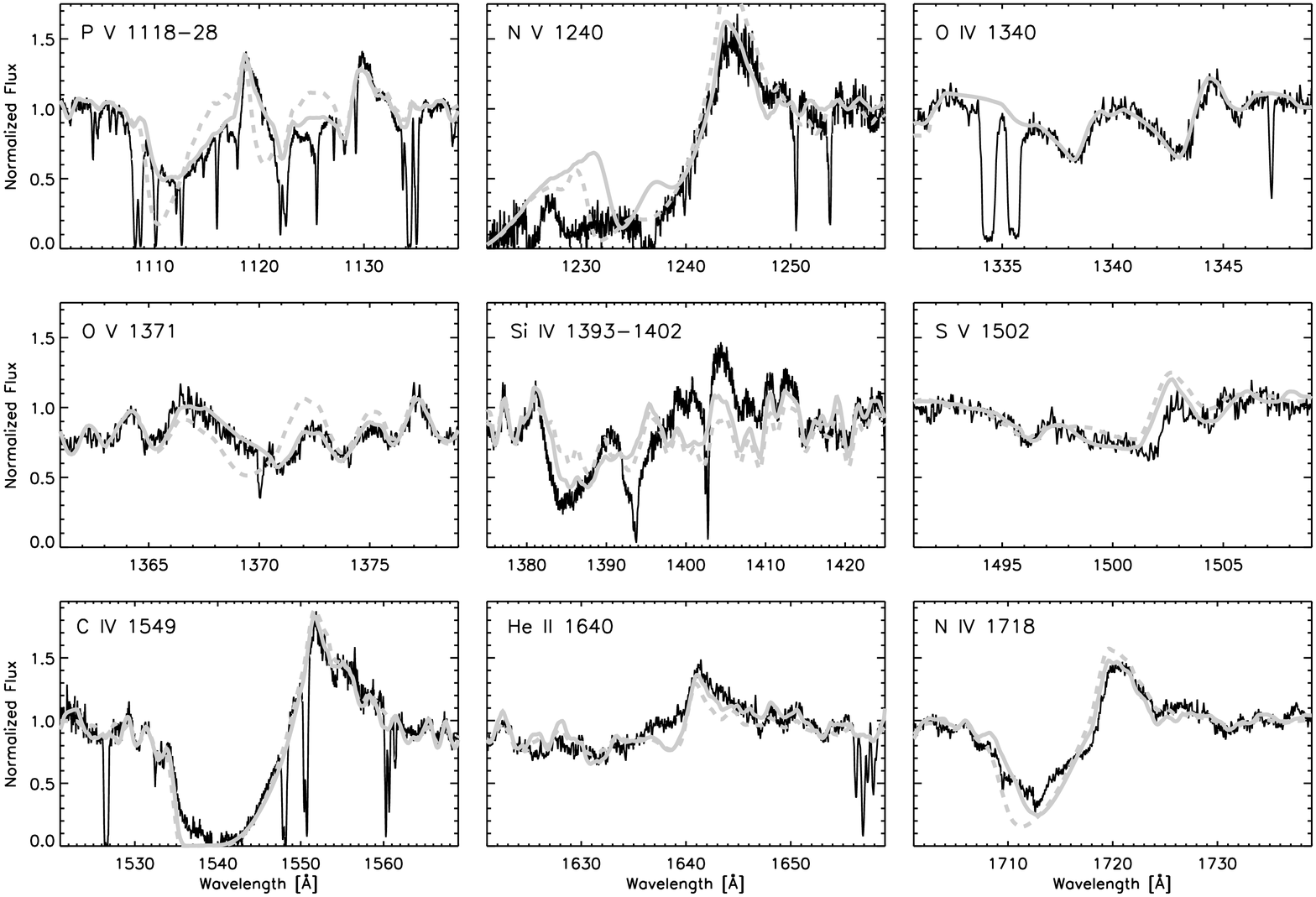}}
      \caption[12cm]{Best fit to \hdo\ wind-sensitive lines, obtained with clumped 
      (full grey line) and homogeneous (dashed grey line) models respectively. The clumped model has
      $f_{\infty}$ = 0.04, $\beta = 0.8$, and \mdot\ = 1.8\,10$^{-6}$\,\msolyr, while the homogeneous
      model has \mdot\ = 6.0\,10$^{-6}$\,\msolyr. The mass loss rates have been adjusted
      as described in \S\ref{hdo_sect}. Adopted abundances for the homogeneous and the clumped
      wind models are discussed in the text (see also Table~\ref{TabRes})}.
      The short wavelength side of \nv\lb1240 is
      affected by interstellar \lyalpha\ absorption (see Table~1 for n(\hi)). Notice in
      particular the excellent fit to both lines of the \pv\  doublet achieved with
      the clumped wind model.
         \label{figtwo}
   \end{figure*}

A detailed examination of the complete FUV spectrum
reveals that only a few lines are influenced by the stellar wind,
even for a supergiant with such an early spectral type and strong
wind (\mdot\ $\approx$ few $10^{-6}$\,\msolyr). In particular, these
lines include a few strong Fe lines whose cores are filled a
little by wind emission. For most other lines, we have a very good
consistency between \tlusty\ and \cmfgen\ model spectra. This
agreement demonstrates that we may use NLTE photospheric models to derive
reliable stellar parameters and abundances, even for Galactic O supergiants 
with relatively dense winds. 
Furthermore, we stress that we had to use large (\xit\ = 15\,\kms)
microturbulent velocities in CMFGEN wind models too. Despite an early comment
by \cite{kudritzki92}, this demonstrates that the need for
microturbulence is not restricted to 
hydrostatic models that neglect the atmospheric velocity field.
We display in Fig.~\ref{figone} the best \cmfgen\ model fit
to some of the lines discussed above, which are only weakly sensitive
to the wind properties.  

We now turn to the analysis of the wind properties of \hdo.
The parameters of the $\beta$-type velocity law are derived from the saturated
\civ\lb\lb1548-1550 doublet and from the \niv\lb1718 P~Cygni profile.
A good match is achieved for $\beta$ = 0.8 and \vinf\ = 2300\,\kms.
We have used an outward increasing microturbulence velocity from the
photospheric value (15\,\kms) to a maximum of 200\,\kms.

To derive the mass loss rate, we started with an estimate
based on \halpha\ modeling, \mdot\ = 1.4\,10$^{-5}$\,\msolyr\ 
\citep{markova04}. Such a high value can be excluded from the FUV spectrum,
because a wind model assuming this value predicts
much stronger P~Cygni profiles than those observed.
Furthermore, this model also produces asymmetric \feiv\ and \fev\ lines 
with a blue extended absorption that is not observed.
\cite{garcia04} derived such a high mass loss rate value from the UV 
wind lines. However, from their Fig.~7, we see that their model predicts \ciii\lb1176 and
\oiv\lb1340 emissions that are not observed. Generally, all other wind lines are also
predicted too strong by their model.
We then built a series of wind models, decreasing \mdot\  in several steps.
Using homogeneous wind models, we have not been able to achieve a good match
to most observed P~Cygni profiles and to other wind sensitive lines like
\oiv\lb\lb1338-1343 and \heii\lb1640.

Therefore, we switched to clumped wind models with the presumption that
the high mass loss rate derived by \cite{markova04} was the result of
the effect of clumps on \halpha\ line formation 
\citep[as already suggested by][]{repolust04}. We searched thus for 
a combination of \mdot\  and $f_\infty$, keeping \mdot/$\sqrt {f_\infty} \approx$
1.4\,10$^{-5}$\,\msolyr. The best match to the FUV lines was
obtained for \mdot\ = 1.8\,10$^{-6}$\,\msolyr and $f_\infty = 0.04$
(see Fig.~\ref{figtwo}). The fit to \pv\lb\lb1118-1128 and to
\ov\lb1371 line profiles are the most striking improvements achieved
with the clumped wind model. The agreement with the observed profile
gets also better for \niv\lb1718, while there are little profile changes
for the \civ\ resonance doublet or \heii\lb1640 compared to profiles
predicted by a homogeneous wind model. The
sensitivity of \pv\ lines to clumping was first established by
\cite{crowther02} and \cite{hillier03}. 
We defer to \S\ref{disc_sect} a discussion of
the physical implications of clumping on line formation.

 \begin{figure*}[]
   \centering
   \rotatebox{0}{\includegraphics[width=18cm]{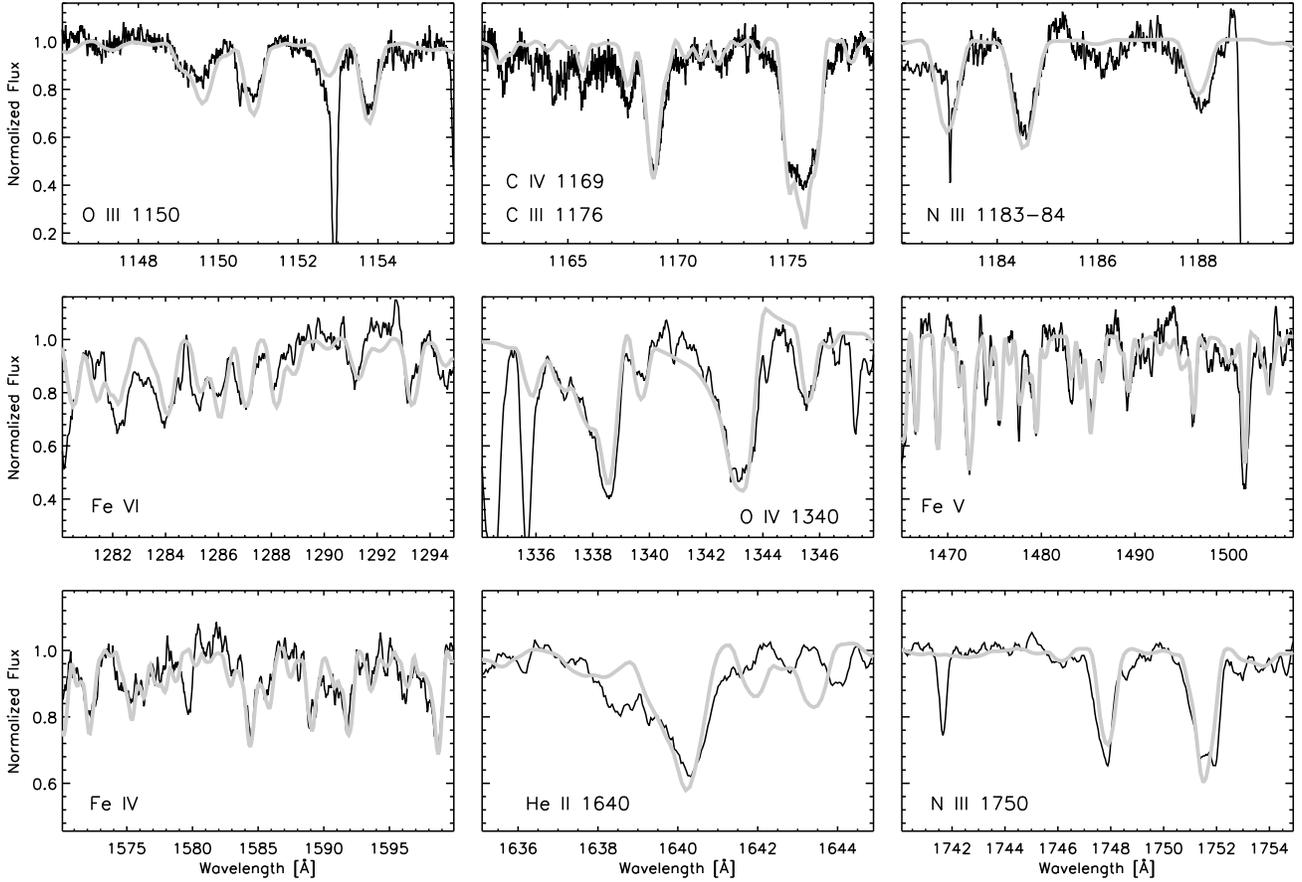}}
      \caption[12cm]{Best fit to photospheric lines used to derive \hdn\ 
      stellar parameters (\teff = 43,500\,K, \logg\ = 4.0, \xit\ = 15\,\kms).
      The adopted abundances are listed in Table~\ref{TabRes}.}
         \label{figthree}
   \end{figure*}

Other wind line profiles, such as \ovi\lb1036 and \nv\lb1240 are strongly
influenced by X-ray wind emission (see \S\ref{discus_2} and Fig.~\ref{figseven}),
and provide little additional constraints on the mass loss rate or on the
clumping properties of the wind of \hdo. P~Cygni profiles of lower ions, like
\ciii\lb977 or \niii\lb991, are severely blended
with \htwo\ lines that hamper an accurate determination of
\mdot\ from these lines.


\subsection{\hdn\ - {\rm O4~V((f))}}
\label{hdn_sect}

The \iue\ spectrum of \hdn\ exhibits conspicuous lines of
\fevi\ (1250-1350\,\AA), \fev\ (1300-1550\,\AA) and
\feiv\ (1550-1700\,\AA), as well as \oiv\lb\lb1338-1343,
\ciii\lb\lb1426-1428 and \sv\lb1502. The \fuse\
spectrum shows \civ\lb1169 and \ciii\lb1176
lines, as well as \oiii\lb\lb1150-1154, \siv\lb\lb1073-1099.
The lines are relatively narrow, implying an apparent
rotational velocity, \vsini\ = 80\,\kms, and resulting in
less severe blending problems from iron lines.

Using the ionization balance between these
ions, we derived a best overall match with a \tlusty\ model
having \teff\ = 43,500\,K and \logg\ = 4.0. A realistic
estimate of the uncertainty on the \teff\  determination
is about $\pm$ 1500\,K. On the other
hand, the sensitivity of the FUV spectrum to \logg\ is
small. We can, however, firmly exclude models with
gravities differing by $\pm$~0.25\,dex in \logg, based on
the comparison of theoretical spectral energy
distributions with the \iue\ spectrum corrected
for reddening (assuming $E(B-V) = 0.42$, Table~\ref{tabone}).
The best overall agreement is obtained for models with \logg\ =
3.9 $\pm$ 0.1. 
The relation between spectral type and surface gravity
for luminosity class~V stars \citep{markova04} yields \logg = 3.9
in agreement with our estimate within the error bars.
We have finally adopted \logg = 4.0,
see Table~\ref{TabRes}.

\cite{garcia04} obtained a markedly lower value, \teff = 39,000$\pm$2,000\,K.
They established  upper and lower temperature limits from the absence of
\pv\lb1118, 1128 and \ov\lb1371. In particular, the weakness of the \ov\
line constrains \teff$<$40,000\,K. 
Homogeneous wind models
predict a very strong \ov\ feature at temperatures higher than 40\,kK,
thus forcing an artificially low \teff\ to be adopted.
Garcia~\& Bianchi methodology yields temperatures that are systematically
lower than all other recent analyses of optical and UV spectra of O stars
(see  \cite{heap05} and \cite{martins05} who
provide a detailed comparison of these studies and who are led to exclude
Garcia~\& Bianchi results).
We believe that their methodology underestimates temperatures
because of the reliance on wind lines that are sensitive to the wind
structure and abundances as well as to \teff.

 \begin{figure*}[]
   \rotatebox{0}{\includegraphics[width=12cm]{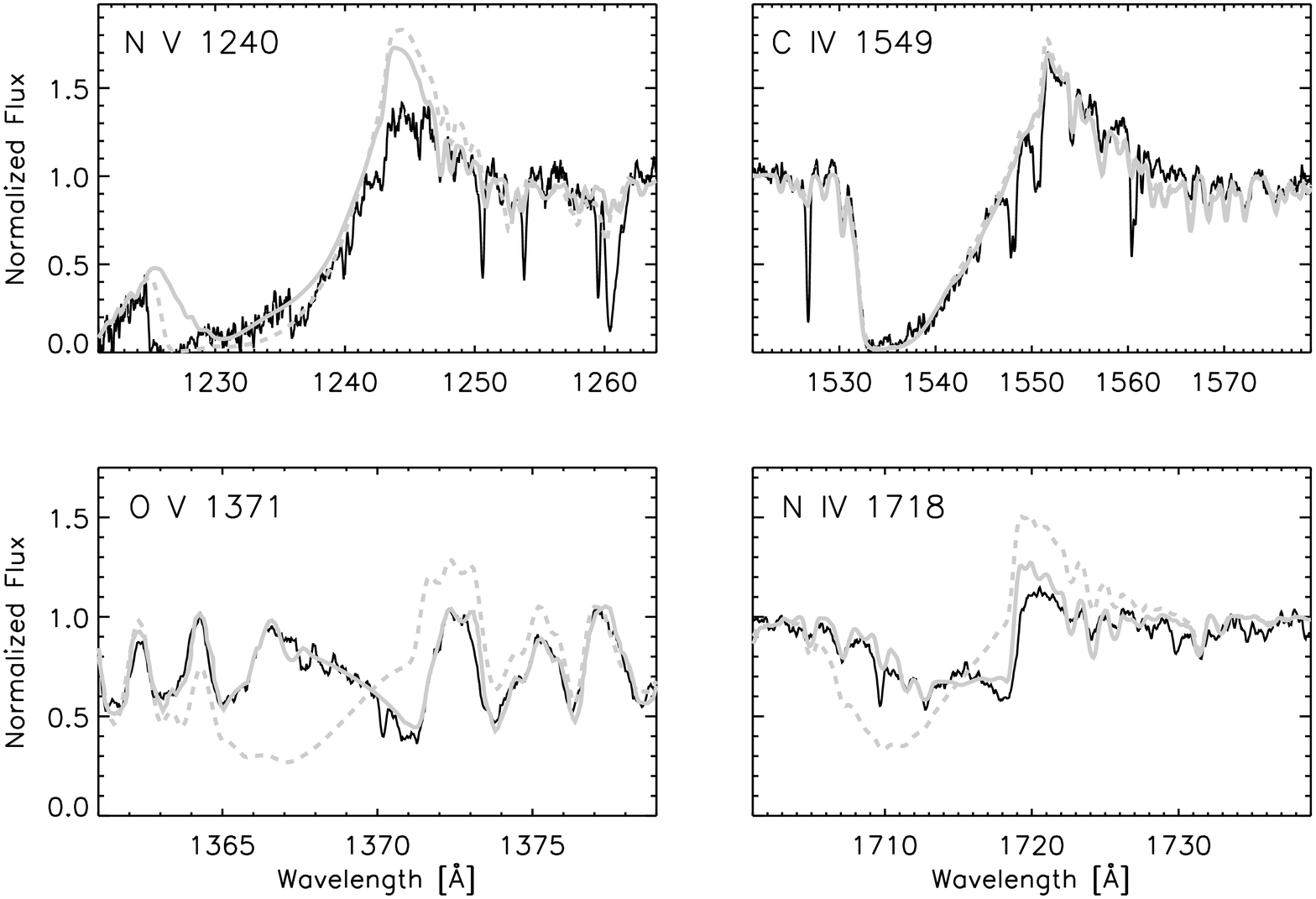}}
      \caption{~Best fit to \hdn\ wind-sensitive lines, obtained with clumped 
      (full grey line) and homogeneous (dashed grey line) models respectively. The parameters of
      the clumped model are
      $f_{\infty}$ = 0.02 and \mdot\ = 2.5\,10$^{-7}$\,\msolyr, while the homogeneous
      model has \mdot\ = 1.8\,10$^{-6}$\,\msolyr. The blue side of \nv\lb1240 is
      affected by IS \lyalpha\ absorption (see Table 1 for n(\hi)).
      The adopted abundances are listed in Table~\ref{TabRes}.}
         \label{figfour}
   \end{figure*}

We adopted the absolute magnitude, $M_{\rm v} = -5.5$, from \cite{howarth89},
and derived the stellar luminosity and radius using a bolometric correction
from \cite{lanz03}. The stellar mass then follows from \logg\ and \rstar.

The microturbulent velocity has been derived from \oiv,
\feiv, \fev, \fevi, and \sv\ lines. The
best match is obtained for \xit\ = 15\,\kms, assuming solar
abundances for these species, consistently with the expected
evolutionary status of \hdn.

We started the analysis using a normal helium abundance, $y = 0.1$.
Subsequently, we slightly decreased it to $y$ = 0.09 to improve
the match to \heii\lb1085 and \heii\lb1640. Admittedly,
this value should be considered as a provisional estimate until an
analysis of the optical spectrum becomes available. Yet, there is
no evidence of helium enrichment at the surface of \hdn.

The carbon abundance was derived from the photospheric lines,
\civ\lb1169, \ciii\lb1176, and
\ciii\lb1426-1428. Using a \tlusty\ model with the stellar parameters
previously derived, we found that C/C$_\odot$ = 0.5 matches well
these carbon lines.

Like \hdo\, we constrained the nitrogen abundance from
\niii\lb\lb1182-1184 and \niii\lb\lb1748-1752. We
found that an enhancement of a factor of 4 with respect to the
solar value is required to match the observed lines. Note
that wind lines such as \niv\lb1718 further support this
determination since no match can be achieved without a
significant nitrogen enrichment. Our result is also consistent
with a line blend, \nii\lb\lb1084-1086, although the presence of
superimposed IS lines makes the determination of the photospheric
contribution difficult.

To measure the oxygen abundance, we relied on \oiii\lb\lb1150-1154
photospheric lines, which are clearly seen in the
\fuse\ spectrum. A very good fit to this triplet is obtained for
O/O$_\odot$ = 0.9, which also results in a good fit of
\oiv\lb\lb1338-1343. Fig.~\ref{figthree} displays the best model fit achieved 
with \cmfgen\ to the photospheric lines discussed in this section. 

The wind velocity law has been determined on the saturated P~Cygni
profile of \civ\lb\lb1548-1550. The terminal velocity and
maximum turbulent velocity have been measured on the blue-side of
the absorption component; we have adopted a terminal wind velocity,
\vinf\ = 3000\,\kms, and a maximum turbulent velocity in the wind
of 250\,\kms. The P~Cygni profile of \civ\lb\lb1548-1550 was
preferred to \nv\lb1240 because the latter is affected by a
significant contamination due to IS \lyalpha\ and because of its
well-known sensitivity to X-rays.
The wind acceleration parameter, $\beta$, and the
mass loss rate have been inferred from the P~Cygni profiles of
\civ\lb\lb1548-1550 and \niv\lb1718 together
with other wind sensitive lines like \oiv\lb\lb1338-1343 and
\heii\lb1640. For the adopted abundances, we found that
a homogeneous wind model with
$\beta$ = 0.8 and \mdot\ = 1.8\,10$^{-6}$\,\msolyr\ reproduces
well these lines, although the match to \niv\lb1718 remains poor.

On the other hand, the \ov\lb1371 line profile predicted by homogeneous
wind models is stronger than observed. This behavior has been found
in all previous studies of O-type stars that show \ov\lb1371.
More specifically, the wind models predict too
strong absorption at moderate velocities, too little absorption at
low velocities, and a too strong redward emission. Homogeneous
wind models thus fail to reproduce this line as in our earlier study
of main-sequence O stars in the SMC \citep{bouret03}. Therefore,
we have calculated several models with different clump volume
filling factors $f_\infty$, adjusting the mass loss
rate so as to maintain a good fit to \civ\ and \niv\ 
P~Cygni profiles. Fig.~\ref{figfour} shows that we can achieve a
good match of \ov\lb1371 when assuming a very small
filling factor, $f_\infty=0.02$, with a mass loss rate, \mdot =
2.5\,10$^{-7}$\,\msolyr. A larger filling factor, $f_\infty=0.1$,
and a lower oxygen abundance by a factor of 4 also improves over
the case of a homogeneous wind model, but does not fit the
observed profile as well as a model with the smaller filling factor.
Larger volume filling factors may seem more reasonable or, at least,
they are more in line with earlier studies of Wolf-Rayet clumped winds
\citep{hamann98} and of SMC O stars \citep{bouret03}. However, other
oxygen lines, like \oiii\lb\lb1150-1154 and \oiv\lb1338-1343, are no
longer fitted with the lower oxygen abundance. These lines are indeed 
sensitive to changes in the adopted oxygen abundance since they are
unsaturated. There is no compelling reason for adopting an oxygen abundance
lower by a factor of 4 relative to the solar value, especially since carbon
is not depleted as well. Carbon would indeed be expected to show first
a depletion typical of CNO-cycle processed material. We have therefore
adopted the clumped model with the very small volume filling factor.

The steep transition between the absorption and
emission components indicates that clumps start forming at low
velocities, $v_{\rm cl}\approx 30$\,\kms, just above the sonic point. 
At this point, we discovered that \niv\lb1718 behaves similarly to
\ov\lb1371, and is thus sensitive to wind clumping too.
The fit to \niv\lb1718 is improved by using
clumped wind models, especially at the transition between the
absorption and emission components. The extended blue absorption
is affected similarly to \ov\lb1371. The nitrogen abundance has been
kept to the value derived from photospheric lines.
Finally, we stress that we have
achieved a good match of both the \niv\ and the \ov\  lines using the
same low filling factor for the clumps, which thus provides convincing
support for the highly clumped model. The profile of other lines
(\civ\ and \nv) is not directly affected by clumping.


\section{Discussion}
\label{disc_sect}

Table~\ref{TabRes} summarizes the results of our
analysis. Conservative estimates of the uncertainties are $\pm$5\% on
\teff, $\pm$0.1\,dex on $\log g$, $\pm$2 to $\pm$5\,\kms\ on
\xit\ (for \hdo\ and \hdn, respectively), and $\pm$0.1 to $\pm$0.2\,dex
on chemical abundances. Abundances are listed as number densities and
refer to values adopted in clumped wind models (lower abundances might be
preferred with homogeneous wind models in a few instances, see \S4).
The mass loss rates are derived with an accuracy of 10-20\,\% from the fitting
process, though this does not include systematic errors.
We discuss now the implications of our results, and compare them to
other observational and theoretical studies.

\subsection{Evidence of clumping from the FUV spectrum}
\label{discus_1}

>From the foregoing analysis, we conclude that \pv\lb\lb1118-1128 and
\ov\lb1371 provide the best diagnostics of clumping in the wind of O stars,
supporting the seminal studies of \cite{crowther02}, \cite{hillier03} and \cite{bouret03},
and extending them to the case of Galactic stars. In attempting to reproduce the
\ov\lb1371 line profile, Bouret et~al. explored a number of alternative
explanations besides clumping. They emphasized that the key point
consists in predicting the correct \ov/\oiv\ ionization structure in the
vicinity of the sonic point. To alter ionization, they considered X-ray
emission from shocks in the wind ($L_X/L_{\rm bol}=2\,10^{-7}$), adiabatic
cooling, larger model atoms, and alternate recombination rates \citep{nahar99}.
All these changes resulted in essentially no difference in the model
spectrum, and wind clumping remained as the only viable explanation.

\begin{table}
\centering \caption{Stellar parameters, wind properties, and chemical abundances
for \hdo\  and \hdn. The listed abundances correspond to values derived from clumped
models (see \S4.1).}
\label{TabRes} 
\begin{tabular}{lcc} \hline \hline
Star                & \hdo                & \hdn        \\ \hline

Spectral Type       & O4 If+              & O4 V((f))  \\         
\teff\  [K]         & 39000               & 43500     \\          
$\log g$ (cgs)      & 3.6                 & 4.0         \\        
$L$ [$L_\odot$]     & 7.9\,$\times\,10^5$ & 4.6\,$\times\,10^5$ \\
$\xi_{\rm t}$ [\kms]& 15                  & 15       \\           
$v \sin i$ [\kms]   & 160                 & 80     \\ \hline             
Homogeneous winds   &  &         \\ \hline           
\mdot\ [\msolyr]    & 6.0\,\evi           & 1.8\,\evi   \\       
\vinf\ [\kms]       & 2300                & 3000       \\         
$\beta$             & 0.8                 & 1.0       \\  \hline          
Clumped winds       &  &            \\ \hline                     
\mdot\ [\msolyr]    & 1.8\,\evi           & 2.5\,\evii    \\    
$f_\infty$          & 0.04                &  0.02      \\  \hline           
Surface abundances$^{\mathrm{a}}$  &  &    \\ \hline      
$y$ (He/H)          & 0.2                 &  0.09         \\       
C/C$_{\odot}$       & 0.05                &  0.5           \\    
N/N$_{\odot}$       & 4.0                 &  4.0           \\     
O/O$_{\odot}$       & 0.1                 &  0.9           \\     
Si/Si$_{\odot}$     & 1.0                 &  1.0           \\     
P/P$_{\odot}$       & 0.5                 &  1.0            \\ 
S/S$_{\odot}$       & 0.9                 &  1.0            \\
Fe/Fe$_{\odot}$     & 1.0                 &  1.0            \\    
\hline       
N/C$^{\mathrm{b}}$  &  80                 &  8      \\           
\hline
\end{tabular}
\begin{list}{}{}
\item[$^{\mathrm{a}}$] Solar abundances from \citet{grevesse98}.
\item[$^{\mathrm{b}}$] Abundance ratio relative to the solar ratio, N/C $\approx$ 0.25.
\end{list}
\end{table}

The case of \pv\  is different. \cite{crowther02} and \cite{hillier03}
showed that an effect on the line profile similar to clumping could be
obtained by drastically reducing the phosphorus abundance because
\pv\  reaches its maximum and is the dominant ion throughout 
the wind of late O supergiants. \cite{hillier03}
achieved a good match to the observed \pv\  doublet in the SMC O7Iaf+ supergiant
AV~83 with homogeneous wind models using a low abundance P/P$_{\odot} = 0.04$
(compared to the adopted overall SMC metallicity, Z/Z$_{\odot} = 0.2$).
While there are remaining uncertainties relative to the phosphorus abundance baseline
and the chemical evolution of the Magellanic Clouds leaving thus some leeway
toward a definitive conclusion, we note that 
\cite{pauldrach94, pauldrach01} similarly
claimed that two Galactic supergiants, namely \zetap\ (O4~I(n)f) and
$\alpha$~Cam (O9.5~Ia) must have a very low
phosphorus abundance based on their analysis of \copernicus\ data. In the case of
Galactic stars, a low phosphorus abundance seems a much less likely explanation
of the weak \pv\  resonance lines, and we thus speculate that Pauldrach et al.'s
results actually indicate that \zetap\  and $\alpha$~Cam also possess
a highly-clumped wind. Our study shows that clumping {\em combined} with
a slightly reduced phosphorus abundance is required to match the \pv\ resonance
doublet in \hdo, while still fitting well the rest of the spectrum.
\cite{hillier03} also achieved an excellent match of the full spectrum
(from FUV to the optical) of AV~83 using $f_\infty$ = 0.1 and P/P$_{\odot}$ = 0.08,
but the evidence for a lower phosphorus abundance remains relatively weak because of
the possibility of adopting a lower clump filling factor coupled to a higher abundance.
These results, therefore, only hint at a depletion of P in O supergiants with
respect to the other metal abundances.

 \begin{figure}[]
\hspace{-0.6cm}
   \rotatebox{0}{\includegraphics[width=9.8cm]{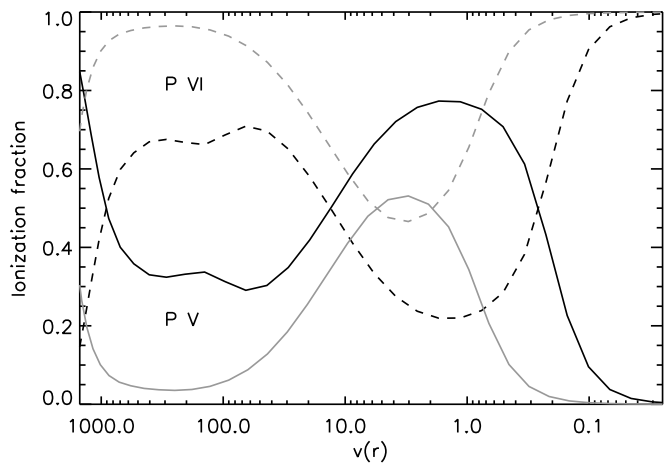}}
      \caption[12cm]{Phosphorus ionization fractions 
      in clumped (black lines) and homogeneous (grey lines) wind models of HD 190429A.
      The \piv\ fraction remains smaller than 1\% throughout the wind
      (max. at $v(r)\approx~1$\,\kms).
      Notice that the velocity scale is shifted toward higher velocities for the
      homogeneous wind model because of a higher mass loss rate (the maximum of
      \pv\  occurs at similar densities in the two models), and that the adopted
      phosphorus abundances are different in the two models (P/P$_{\odot}$ = 0.1 and
      0.5, for the homogeneous and clumped models, see \S4.1).}
         \label{figfive}
   \end{figure}

In addition to \ov\  and \pv, we finally found that \niv\lb1718 is also
an indicator of wind clumping. The \niv\  line profile exhibits the same
behavior as \ov\lb1371. The two lines correspond to the same transition,
2s($^2$S)2p~$^1$P$^0$ -- 2p$^2$~$^1$D, in an isoelectronic sequence. 
The \ov\  and the \niv\  lines can be fitted with a clumped wind model
having the same clumping parameters.

For both stars, we found small volume filling factors, {$f_\infty = 0.04$} and
0.02, for \hdo\  and \hdn, respectively. Although these small values may
seem problematical, they are supported by other recent results about wind clumping.
In the analysis of 25 O stars in the LMC, \cite{massa03} found that the \pv\  ionization
fraction never exceeds 0.20, although it is expected to be the dominant stage
in some of the stars. They argued that this implies either that the calculated
mass loss rates or the phosphorus abundance are too large, or that the winds are
strongly clumped ($f_\infty < 0.05$). The first evidences of wind clumping were
found in WR stars for which small filling factors ($f_\infty \approx 0.1$) were derived
too \citep{hamann98, hillier99, crowther02d}. \cite{kurosawa02} and \cite{schild04} derived
even lower values, $f_\infty \approx 0.05 - 0.075$, for the WR stars V444~Cyg and
$\gamma^2$~Vel. Hydrodynamical 1-D calculations of line-driven wind instabilities support
low volume filling factors \citep[$f_\infty < 0.1$;][]{runacres02}, but initial results of
a restricted 2-D approach indicate larger values \citep[$f_\infty \approx 0.2$;][]{dessart03}.

\subsection{Effect of clumping on wind ionization}
\label{discus_2} 

We now examine the reasons why these line profiles are modified by clumping.
Essentially, the presence of optically thick clumps
separated by transparent voids has two major consequences: {\it (i)} the wind
over-densities strengthen the emission of density-sensitive lines; {\it (ii)} the
number of recombinations increases because of the higher density in the clumps,
thus reducing ionization. The competition between these two effects may
either increase or decrease emission.
Because of the radial dependence of the volume filling factor,
lines of different strength behave in different ways. This is shown
in Fig.~\ref{figtwo} and Fig.~\ref{figfour} where profiles for clumped and homogeneous
models are compared for \hdo\ and  \hdn.

 \begin{figure}[]
\hspace{-0.6cm}
   \rotatebox{0}{\includegraphics[width=9.8cm]{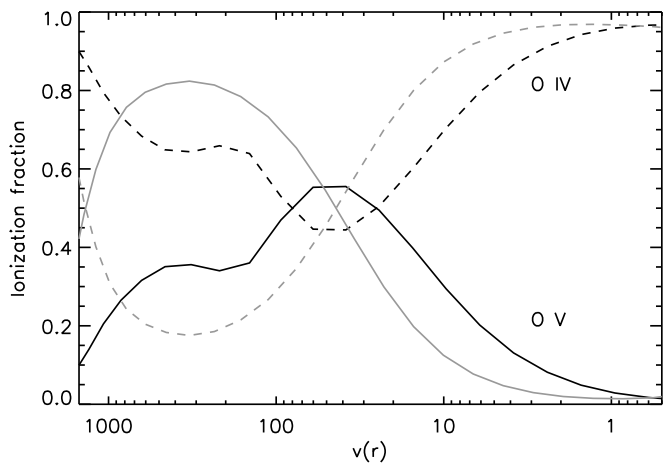}}
      \caption[12cm]{Oxygen ionization fractions 
      in clumped (black lines) and homogeneous (grey lines) wind models of HD 96715.
      Notice that the velocity scale is shifted toward higher velocities for the
      homogeneous wind model because of a higher mass loss rate.}
         \label{figsix}
   \end{figure}

In O7 supergiants, \pv\  is the dominant ionization stage of phosphorus throughout
the wind \citep{hillier03}. However, \hdo\  being a hotter supergiant, \pv\  (in the photosphere)
and \pvi\  (in the wind) are now the main ionization stages. The \piv\  fraction always remains
small (less than 1\%). Figure~\ref{figfive} illustrates the effect of clumps in the stellar
wind: we find a lower ionization, \pvi/\pv, compared to the homogeneous wind model,
demonstrating the increased recombination from \pvi\  to \pv\  in the clumps.
In \hdo, the \pv\lb\lb1118-1128 resonance lines reflect the factor of 2 between
the $f$-values of the two lines, thus indicating that these lines are unsaturated.
The interpretation of the strength of these lines is, however, not straightforward,
that is, it depends not only on the phosphorus density (a product of the mass loss
rate and the phosphorus abundance), but it also depends on the detail of the wind
ionization. The dependence with the clumping parameter, $f_\infty$, is indirect
through the requirement that the ratio \mdot/$\sqrt {f_\infty}$ has to remain constant for
keeping a good match to other lines, such as H$\alpha$ and \civ\lb1550.
A low phosphorus abundance (P/P$_\odot=0.1$) must be adopted with a homogeneous
wind model, thus implying significant clumping (low $f_\infty$, hence low \mdot) is necessary
to increase the phosphorus abundance to a near-solar value. \cite{massa03} followed
the same argument to explain the systematically low \pv\  ionization fractions ($<$20\%)
that they derived for 25 O stars in the LMC. We stress that our clumped wind model matches
the {\em two} \pv\  resonance lines very well, providing a strong evidence that the
wind ionization is correct and that there is no particular issue such as covering factors
with our treatment of clumps.

The oxygen lines, however, respond in a different way to clumping.
They are affected by clumping via a lower ionization in the wind -- see Fig.~\ref{figsix}
for \hdn. Notice that ionization in fact is not lower in the photosphere as Fig.~\ref{figsix}
would suggest: there is a shift in velocity space because of the higher mass loss rate
of the homogeneous wind model. At similar photospheric densities, the ionization is the
same. On the other hand, \oiv\  is the dominant ion throughout the wind of \hdo.
The homogeneous model predicts an increase of \ov\  at moderate velocities (over 30\%
at few 100's \kms), but increased
recombination due to the presence of clumps keeps the \ov\  fraction below a few percent
in the clumped wind. The lower \ov\  fraction in the clumped wind models weakens the
blue-shifted line absorption substantially.
The wind ionization is sensitive to the clumping parameter, ${f_\infty}$,
and \ov\lb1371 is the best indicator of clumping in the wind of \hdn, similarly to
the SMC O4 stars \citep{bouret03}.

Besides the \ov\  and \pv\  lines, we found that \niv\lb1718 is also sensitive
to clumping. It is a special interest because clumping decreases \nv/\niv\ 
ionization in a way quite similar to \ov/\oiv. \niv\  is the dominant ion throughout
the wind of \hdo, while it is the dominant ion only in the photosphere of \hdn\ 
(\nv\  is the dominant stage in the wind). The \niv\lb1718 responds
to clumping like \ov\lb1371. The absorption
component of \niv\lb1718 is globally weaker in the clumped model, with the
exception of velocities between $\approx$ 15\,\kms\ and 650\,\kms\ 
(Fig.~\ref{figfour}). The direct mapping of the velocity law on the absorption
component implies that these velocities correspond to the region where \nv/\niv\ 
ionization is lower in the clumped model. In the outer wind, at higher velocities,
\nv\  recombines into \niv\  in the homogeneous and clumped models of \hdn\  because
of lower temperatures. The predicted profile at these high velocities ($v > 650$\,\kms)
then reflects the higher mass loss rate adopted for the homogeneous wind model.

 \begin{figure}[]
\hspace{-0.6cm}
   \rotatebox{0}{\includegraphics[width=9.8cm]{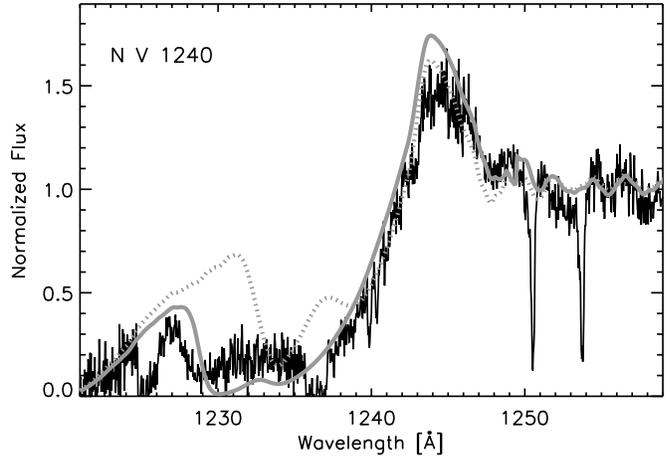}}
      \caption[12cm]{Effect of X-ray wind emission on the predicted \nv\lb1240 line profile
      of clumped wind models of \hdo\  (full line: \logLX\ $= -7.1$; dotted line: no X-rays).
      All the other lines sensitive to wind clumping and shown in Fig.~\ref{figtwo}
      are not affected by X-ray emission.}
         \label{figseven}
   \end{figure}

Several other lines are displayed in Fig.~\ref{figtwo}, showing smaller changes
because of clumping. For instance, \sv\lb1502 shows a slightly stronger blueshifted
wing because of increased recombination from \svi\  to \sv. On the other hand, the
\civ\ resonance doublet is only marginally sensitive to clumping
because these lines are saturated.

X-ray emission in the wind also changes the wind ionization structure, increasing the
populations of ``superions'' like \nv, \ovi, and \svi. We found that the profiles
of \nv\lb1240 and \ovi\lb1036 are substantially changed by X-ray emission (see Fig.~\ref{figseven};
the \ovi\  lines are not shown because of strong interstellar H$_2$ absorption).
All other FUV lines shown in Fig.~\ref{figtwo}, especially those sensitive to wind clumping,
remain unaffected. In the outer wind of \hdo\  ($v(r) > 500$\,\kms), the 
\nv\ is increased
by a factor of several up to several orders of magnitude at terminal velocity by the
X-ray wind-shock emission compared to predictions neglecting X-rays. In the inner wind
($v(r) < 500$\,\kms), \nv\ increase is more modest ($+$20\% compared to predictions without
X-rays). 
The \nv\ fraction reaches about 10\% in the inner wind and always remains larger than
1\% in the outer wind while it precipitously drops in the model without X-rays.
This change in the wind ionization structure results in the marked line profile change
displayed in Fig.~\ref{figseven}.

Finally, we need to stress here that clumping is not a simple numerical trick introduced
to correct the ionization balance of a single species (e.g., \ov/\oiv). We showed
here that our clumped models change in a consistent way the ionization balance
of different species, with different ionization potentials, and
affect the corresponding lines differently as they form in different parts
of the wind and respond differently to the nature of clumping in the wind.
We therefore conclude that our description of wind clumping, albeit simple, must
be basically correct because of our ability to fit lines of different ions and species
using the {\em same} clumping parameters. The clumped wind models therefore provide
a sound basis for determining improved mass loss rates.

\subsection{Clumping, \halpha, and the mass loss rate}
\label{discus_3}

We discuss now the implication of wind clumping
on the derived mass loss rates, comparing our results based on the analysis of
the FUV spectrum with determinations based on \halpha\  and thermal radio emission.
Such a comparison is essential because the quantity that is actually measured
from the FUV lines is the product of the mass loss rate by the appropriate
ionization fraction, and we just argued that clumping results in
changing the mean ionization in the wind. We believe that \cmfgen\  models provide
a reliable description of the ionization structure of O star winds thanks
to the detailed, unified treatment of NLTE metal line blanketing in the photosphere
and in the wind. Yet, a comparison with results derived from \halpha\  is crucial
since the latter are much less sensitive to many factors influencing the wind
ionization. Moreover, the FUV lines behave linearly with the wind density
while \halpha\  and the radio emission have a quadratic dependence.

\halpha\ is a recombination line, and is sensitive to the densest regions of the wind.
Therefore, this line is expected to constrain clumping deep in the wind.
On the other hand, thermal emission in the radio probes the outer
regions of the wind. \cite{kudritzki00} argued that \halpha\  should not
be sensitive to wind clumping contrary to the submillimeter/radio emission
because they expected clumps to form far out in the wind. This is
contrary to our findings that clumps already form close to the sonic point.

 \begin{figure}[]
   \centering
   \rotatebox{90}{\includegraphics[width=6cm]{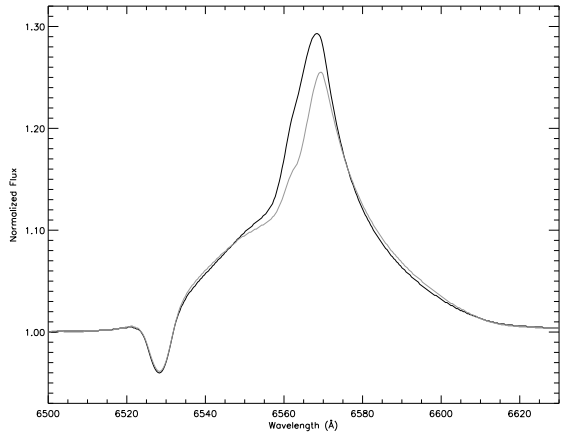}}
      \caption[12cm]{Predicted \halpha\ profiles for \hdo, calculated with clumped 
      (grey line) and homogeneous (black line) wind models. The corresponding
      parameters are listed in Table~4.}
         \label{figeight}
   \end{figure}

Although we have no \halpha\  observations in hand, we can compare our
predictions for homogeneous and clumped wind models with published results
for \hdo. We display in Fig.~\ref{figeight} the predicted \halpha\ 
profiles calculated with the parameters derived from our analysis
(see Table~\ref{TabRes}). We found only little changes between the predicted
profile for the homogeneous and clumped wind models, thus indicating
that \halpha\  observations alone would not exclude one or the other model
(but different mass loss rates would be inferred). Our predicted line profiles are
consistent with Walborn~\& Howarth's (2000; their Fig.~2) published profile.

Table~\ref{TabWind} lists the mass loss rates derived from \halpha\  and radio
observations and compares them to our results. The mass loss rates (\halpha\  and radio)
depend on the adopted distance, proportionally to $d^{3/2}$ (e.g., Scuderi et al.
1998). The values listed in Table~\ref{TabWind} have been corrected of this
dependence. The mass loss rate for a homogeneous
wind derived from the FUV lines is consistent with earlier analyses of
\halpha\  \citep{leitherer88, lamers93}, but \cite{scuderi98} and \cite{markova04}
derived a value twice as large. Unified comoving-frame models were used by Markova et~al.
and in our study, while the other analyses used the Sobolev approximation. 
\cite{scuderi98} also report an increase by a factor of 2
of the \halpha\  equivalent width between 1988 and 1991. They obtained also a
single observation of the free-free emission at 8.45\,GHz. It is therefore
difficult to reach a firm conclusion from this comparison since intrinsic
stellar variability most likely is at the origin of some of the differences.
The main conclusions thus remain that {\it (i)} the profile from the clumped 
wind model
would match the observed \halpha\  profiles as well as the predicted profile
from the homogeneous wind, and {\it (ii)} the derived mass loss rate from the
clumped wind model is a factor of 3 lower.

 \begin{figure}[]
   \centering
   \rotatebox{90}{\includegraphics[width=6cm]{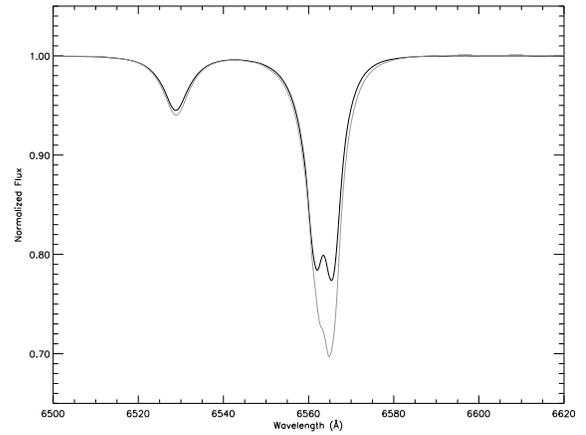}}
      \caption[12cm]{Predicted \halpha\ profiles for \hdn, calculated with clumped 
      (grey line) and homogeneous (black line) wind models. The corresponding
      parameters are listed in Table~4.}
         \label{fignine}
   \end{figure}
   
Finally, hydrogen infrared recombination lines such as \bralpha\ and \brgamma\
may be strongly affected by clumping in the wind, because of their sensitivity
to the square of the local density. As pointed out by \cite{lenorzer04}, clumping
results in increased recombination rates inside the clumps 
($\langle \rho^2 \rangle > \langle \rho \rangle^2$, assuming an infinite 
contrast of the clumps with the voids). The presence of clumps deep
in the wind would thus explain the present 
discrepancies between observed and theoretical profiles (and EWs) of lines such as
\brgamma\ \citep{lenorzer04}. Indeed, clumping close to the photosphere
(where \brgamma\ forms) would increase the collisional excitation rates
driving the excitation ratios closer to LTE. These recombination lines would thus exhibit
(pure) emission profiles rather than P Cygni profiles \citep[cf.][]{puls96}.
An interesting point noted by \cite{lenorzer04} is that their predicted
\bralpha\ agree fairly well with observations. This suggests that 
the degree of clumping might not be constant throughout the wind,
but rather peaks and then decreases further out in the wind. The location of 
the strongest clumping (expected to depend on the maximum line strength according to
\cite{runacres02}) might then be in the region where \brgamma\ forms.
\bralpha\ which forms over a larger volume (because of its larger oscillator
strength) would then be less affected by clumping. 

\begin{table}
\centering \caption{Mass loss rates and modified wind momenta measured for \hdo\  and \hdn.}
\label{TabWind} 
\begin{tabular}{lcc} \hline \hline
Star                & \hdo                & \hdn        \\
\hline
\multicolumn{3}{c}{Mass Loss Rates [$10^{-6}$\,\msolyr]}            \\
\multicolumn{3}{c}{This study (far-UV)}            \\
Clumped wind                   &  1.8$\pm$0.3   &  0.25$\pm$0.05 \\
Homogeneous wind               &  6.0           &  1.8          \\
\multicolumn{3}{c}{H$\alpha$}            \\
\cite{leitherer88}               & 9.8 & \\
\cite{lamers93}                &  5.4$^{+3.1}_{-2.3}$  &  \\
\cite{scuderi98}               &  11.8$\pm$1.8   &  \\
\cite{markova04}               & 14.2 & \\
\multicolumn{3}{c}{Radio}            \\
\cite{scuderi98}               &  7.2$\pm$0.9   &  \\  \hline
\multicolumn{3}{c}{Modified Wind Momentum}            \\
\multicolumn{3}{c}{$\log$\,(\mdot\,$v_\infty\,\sqrt{R_\star})$}            \\
Clumped wind                   &  29.07   &  28.21 \\
Homogeneous wind               &  29.66   &  29.07 \\
\cite{repolust04}              &  29.54   &  29.11 \\
\cite{vink00}                  &  29.55   &  29.13 \\
\hline
\end{tabular}
\end{table}

Our estimate of \mdot\  for \hdn\ derived from
the homogeneous wind model compares well to mass loss rates for stars
of same spectral type \citep{lamers93, repolust04}. \halpha\ is always observed
in absorption for these stars, which is predicted both for the clumped and
homogeneous wind models. Fig.~\ref{fignine} shows that \halpha\ is
somewhat filled by wind emission for the homogeneous wind model, but it remains
unlikely that either model could be excluded based on \halpha\  observations.
The clumped model yields a mass loss rate that is 7 times lower than the value derived
from the homogeneous wind model.

\subsection{The Wind-momentum Luminosity Relation}
\label{discus_4}

The presence of clumps deep in the wind of O stars has been recently
invoked by \cite{puls03} and \cite{repolust04} to explain their findings of two
distinct Wind-momentum Luminosity Relations (WLR) for stars with \halpha\ in
emission and in absorption, respectively. While the
WLR for the latter objects agrees with \cite{vink00} theoretical predictions, it
is clearly separated from the empirical WLR for emission-line objects by a significant
offset. As shown in \cite{markova04} and \cite{repolust04}, this offset
practically vanishes when the mass loss rates for the emission-line stars are
reduced by a factor of 0.44. Because clumped winds mimic winds with higher mass loss
rates, these authors suggested that the higher empirical WLR for emission-line
objects reflects the influence of clumps in the \halpha\  line formation region.

From our analysis, we conclude that mass loss rates from the homogeneous wind models
are very similar to those predicted by the \cite{vink00} formula using the parameters
derived for \hdo\ and \hdn.
Table~\ref{TabWind} lists Vink et al.'s predicted modified wind momenta 
together with the values derived from our analysis. As might be expected from the
lower mass loss rates derived from the clumped wind models, the wind momenta
calculated from these models are significantly lower than the theoretical
predictions. Our results cast doubt on \cite{repolust04} and \cite{markova04} tentative
suggestion to lower supergiant mass loss rates in order to
reunify the empirical WLRs with the theoretical one. 
We have indeed shown that a correction for clumping is required even for dwarfs that
are expected to show \halpha\  in absorption (or partially filled in by wind emission). 
This, in turn, implies that the mass loss rates should be lowered, leaving thus
a difference between the supergiant and dwarf WLRs. 
Moreover, \cite{repolust04} mention that the blue Balmer lines in  stars with \halpha\ emission
show too much wind emission in their cores, pointing to too high \halpha\  mass loss rates. 
Clumps at the base of the wind might account for this behaviour. We finally note that
the existence of a relatively deep-seated clumped region is compatible with recent,
theoretical results \citep[e.g.][ and references therein]{feldmeier97, owocki99} .
Although the stellar sample is limited, we emphasized that we found evidence of wind
clumping in all early O dwarfs that we have analyzed so far, see \cite{bouret03}.
A unique WLR for supergiants and dwarfs might eventually be empirically derived
with clumped wind models once larger stellar samples have been analyzed.
We found, however, that wind clumping results in revising down
the mass loss rates by a factor larger than the factor of about 2.3
advocated by \cite{repolust04}.

   
\section{Conclusions}
\label{concl_sect}

We have performed a quantitative analysis of the \fuse\  and \iue\   spectra of two
Galactic O4-type  stars, one supergiant (\hdo) and one dwarf (\hdn),
investigating the role of wind clumping and its resulting effect on stellar
parameters and thus extending our recent studies of SMC O stars \citep{bouret03,
hillier03}. Our analysis is based on the NLTE model atmosphere programs
\tlusty\  and \cmfgen, that account for NLTE metal line blanketing and thus
provide a detailed description of the photosphere and wind of O stars. We have
first determined the stellar and wind parameters using homogeneous wind models,
but failed to reproduce several key lines, e.g. \ov\lb1371, like all previous
analyses of O stars. Similarly to the case of MC stars \citep{crowther02, hillier03},
we have to adopt a very low
phosphorus abundance in order to match the strength of the \pv\lb\lb1118-1128
resonance lines with homogeneous wind models. We have then considered clumped
wind models in order to test their ability to improve the fit to the observed
spectra.

The clumped wind models {\em consistently} improve the match to lines of different species,
especially \pv\lb\lb1118-1128, \ov\lb1371 and \niv\lb1718. The fit to strong Fe photospheric
lines (similarly to AV~83, Hillier et~al. 2003) and to \siiv\lb\lb1393-1402
(though still not excellent) is also improved in \hdo. In both stars, we need to adopt
a highly clumped wind model, $f_\infty = 0.04$ (\hdo) and $f_\infty = 0.02$ (\hdn), in
order to match these lines. 
Based on measured
phosphorus ionization fractions, \cite{massa03} argued that mass loss rates of O stars
might be lower by a factor of 5, which is thus consistent with $f_\infty < 0.1$.
In agreement with \cite{hillier03} and \cite{bouret03},
clumping needs to start deep in the wind, just above the sonic point, at velocities
as low as $v_{\rm cl} \approx 30$\,\kms, to reproduce the steep transition between
the emission and absorption components of clump-sensitive lines. This finding also
supports the conjecture of \cite{repolust04} that their mass loss rates based on \halpha\ 
were too high for supergiants due to clumping. We showed that the basic physical reason
behind the better fits is the increased recombination in clumps that lowers the wind
ionization. We argue that our clumped wind models accurately predict the ionization
structure of O-type star winds because of their ability to consistently match lines
of different species and ionization stages. Our results, therefore, provide
a considerably more robust conclusion about the clumped nature of the wind of O stars
than previous studies, and are in agreement with our earlier results on SMC O stars
\citep{hillier03, bouret03}.

The main consequence of wind clumping is that mass loss rates need to be revised down
substantially, here by a factor of 3 (\hdo) to 7 (\hdn). This factor is of the same order
as the correction derived for the SMC O stars \citep{bouret03}. We show that such a drastic
correction remains in good agreement with \halpha\  observations.
Most earlier studies of O stars ignored the effect of deviations from standard winds models,
i.~e. assuming a globally stationary wind with a smooth density/velocity stratification,
to determine the properties of stellar winds. It was argued that ``this standard analysis
yields reliable average models of the stellar wind'' \citep{kudritzki00}, although it
was recognized that such models
are inherently incapable of reproducing the spectral signatures that suggest the existence of 
extensive wind structures.
Although this argument may hold for deriving the wind velocity structure, we have
now demonstrated that this is not the case for determining mass loss rates. The present
work, together with our recent work on SMC O stars, establishes that clumping
likely is a general property of O star winds.
Accounting for clumping will lead to a systematic and significant downward revision
of mass loss rates. Because mass loss is a crucial aspect of massive star evolution,
a revision of mass loss rates accounting for the effect of clumping is urgently
needed. We note in parallel that two analyses of low-luminosity O stars also revealed
mass loss rates much lower than those predicted by the theoretical WLR \citep{bouret03,
martins04}. Our study therefore calls for a fundamental revision in our understanding
of mass loss and of O-type star stellar winds.

Finally, the surface abundance of \hdo\  is in agreement with its advanced evolutionary stage,
in particular showing CNO-cycle processed material (nitrogen overabundance, carbon and
oxygen depletion) at the stellar surface. Similarly to SMC O dwarfs, we find also
enhanced nitrogen at the surface of \hdn, though the carbon depletion is much milder
and oxygen is close to the solar abundance.

\begin{acknowledgements}
We are grateful to Joachim Puls for a detailed and helpful referee's report
which contributed to improve our paper. 
All data used in this paper were extracted from the Multimission Archive at 
the Space Telescope Science Institute (MAST). 
STScI is operated by the Association of Universities for Research in Astronomy, Inc., 
under NASA contract NAS5-26555. Support for MAST for non-HST data is provided by the 
NASA Office of Space Science via grant NAG5-7584 and by other grants and contracts.
J.-C. Bouret acknowledges CNES for financial support. T. Lanz is supported by
a NASA ADP grant (NNG04GC81G); he is grateful for the hospitality and support
of the Laboratoire d'Astrophysique in Marseille when this work was initiated.
D.~J. Hillier gratefully acknowledges support from a NASA-LTSA grant (NAGW-3828).
\end{acknowledgements}

\bibliographystyle{aa}
\bibliography{article}


\end{document}